\documentclass[twocolumn,prl,preprintnumbers,nofootinbib]{revtex4}

\usepackage{graphicx}
\usepackage{bm}

\newcommand{\lsim}{~{}_{\textstyle\sim}^{\textstyle <}~}

\newcommand{\be}{\begin{equation}}
\newcommand{\ee}{\end{equation}}

\newcommand{\bea}{\begin{eqnarray}}
\newcommand{\eea}{\end{eqnarray}}

\begin{document}

\preprint{CALT-68-2554}
\preprint{KRL-MAP-307}

\title{How Magnetic is the Dirac Neutrino? }
\author{Nicole F. Bell}
\author{V. Cirigliano} 
\author{M. J. Ramsey-Musolf}
\author{P. Vogel}
\author{Mark B. Wise}
\affiliation{
California Institute of Technology, 
Pasadena, CA 91125, USA 
} 

\date{December 16, 2005}

\begin{abstract}
 
 We derive model-independent, \lq\lq naturalness" upper bounds on the
 magnetic moments $\mu_\nu$ of Dirac neutrinos generated by physics
 above the scale of electroweak symmetry breaking. In the absence of
 fine-tuning of effective operator coefficients, we find that current
 information on neutrino mass implies that $|\mu_\nu|\lsim 10^{-14}$
 Bohr magnetons. This bound is several orders of magnitude stronger
 than those obtained from analyses of solar and reactor neutrino data
 and astrophysical observations.

\end{abstract}

\pacs{Valid PACS appear here}
\maketitle

With the current emphasis on understanding the pattern of neutrino
mass and mixing and the corresponding implications for cosmology and
astrophysics, it is also of interest to consider the electromagnetic
properties of the neutrino. The leading coupling of the neutrino to
the photon is the magnetic moment, $\mu_\nu$. The chiral symmetry
obeyed by the massless neutrinos of the Standard Model (SM) requires
that $\mu_\nu=0$. Now that we know that $m_\nu\not=0$, however, it is
interesting to ask how large one might expect the neutrino magnetic
moment to be. In the minimally extended SM containing gauge-singlet
right-handed neutrinos, one finds that $\mu_\nu$ is non-vanishing, but
unobservably small: $\mu_\nu\approx 3\times 10^{-19}\mu_B[m_\nu/1\
{\rm eV}]$ \cite{Marciano:1977wx}.

Current experimental limits are several orders of magnitude
larger. Those obtained from laboratory experiments are based on
analyses of the recoiling electron kinetic energy $T$ in
neutrino-electron scattering. The effect of a non-vanishing $\mu_\nu$
will be recognizable only if the corresponding electromagnetic cross
section is comparable in magnitude with the well-understood weak
interaction cross section. The magnitude of $\mu_\nu$ which can be
probed in this way is then given by
\begin{equation}
\label{eq:muexp}
\frac{|\mu_{\nu}^{\rm exp}|}{\mu_B} 
\simeq \frac{G_F \, m_e}{\sqrt{2} \pi \alpha} \sqrt{m_e T}
\sim  10^{-10} \sqrt{\frac{T}{m_e}}   \ \ .
\end{equation}
Considering realistic values of $T$, it would be difficult to reach
sensitivities below $\sim10^{-11} \mu_B$.  The limits derived from
studies of solar and reactor neutrinos are presently somewhat weaker:
$|\mu_\nu |\lsim 1.5 \times 10^{-10}\mu_B$ (solar) \cite{beacom} and
$|\mu_\nu |\lsim 0.9 \times 10^{-10}\mu_B$ (reactor), \cite{MUNU}.

Limits on $\mu_\nu$ can also be derived from bounds on unobserved
energy loss in astrophysical objects. For sufficiently large
$\mu_\nu$, the rate for plasmon decay into $\nu{\bar\nu}$ pairs would
conflict with such bounds. Since plasmons can also decay weakly into
$\nu{\bar\nu}$ pairs, the sensitivity of this probe is
again limited by size of the weak rate, leading to
\cite{Sutherland:1975dr}
\begin{equation}
\frac{|\mu_{\nu}^{\rm astro}|}{\mu_B} 
\simeq \frac{G_F \, m_e }{\sqrt{2}\pi \alpha } \left(\hbar\omega_P\right)\ \ ,
\end{equation} 
where $\omega_P$ is the plasma frequency. Since $(\hbar \omega_P)^2
\ll m_e T$, 
this
bound is stronger than the limit in Eq.
(\ref{eq:muexp}).  Given the appropriate values of $\hbar \omega_P$ it
would be difficult to reach sensitivities better than $10^{-12}
\mu_B$. Indeed, from the analysis performed in Ref. \cite{raffelt},
one obtains $|\mu_\nu|\lsim 3\times 10^{-12}\mu_B$.

In what follows, we show -- in a general and model-independent way --
that a magnetic moment of a Dirac neutrino with magnitude of the same
order, or just below, current limits would be unnaturally large and
would require the existence of fine tuning in order to prevent
unacceptably large contributions to $m_\nu$ {\em via} radiative
corrections\footnote{The idea that SM-forbidden operators might
contribute to $m_\nu$ through loop effects was first proposed in
Ref. \cite{Schechter:1981bd} and recently discussed in
Ref. \cite{Ito:2004sh}.}. 
Although small Dirac neutrino masses imply very small Yukawa couplings, 
they are not inconsistent with observations.
In order to satisfy $m_\nu \alt 1 {\rm eV}$, we argue that a more 
natural scale for $|\mu_\nu|$ would be $\lsim 10^{-14} \mu_B$.

Assuming that $\mu_\nu$ is generated by some physics beyond the SM at
a scale $\Lambda$, its leading contribution to the neutrino mass,
$\delta m_\nu$, scales with $\Lambda$ as 
\begin{equation}
\label{eq:leading} 
\delta m_\nu \sim \frac{\alpha}{32\pi} \frac{\Lambda^2}{m_e}
\frac{\mu_\nu}{\mu_B}\ \ , 
\end{equation} 
where, $\delta m_\nu$ is the contribution to a generic entry in the
$3\times 3$ neutrino mass matrix arising from radiative corrections at
one-loop order. The dependence on $\Lambda^2$ arises from the
quadratic divergence appearing in the renormalization of the dimension
four neutrino mass operator. Although the precise value of the
coefficient on the right side of Eq. (\ref{eq:leading}) can only be
obtained with the use of a specific model, it implies an
order-of-magnitude bound on the size of $\mu_\nu$. For $\Lambda\sim 1$
TeV, requiring that $\delta m_\nu$ not be significantly larger than
one eV implies that $|\mu_\nu| \lsim 10^{-14} \mu_B $. Given
the quadratic dependence on $\Lambda$, this bound becomes considerably
more stringent as the scale of new physics is increased from the scale
of electroweak symmetry breaking, $v\sim 250$ GeV.

The problem of reconciling a large magnetic moment with a small mass
has been recognized in the past, and the quadratic dependence on
$\Lambda$ in Eq.~(\ref{eq:leading}) discussed in, e.g.,
\cite{Barr,Voloshin}.  Possible ways of overcoming this restriction
include imposing a symmetry to enforce $m_\nu=0$
while allowing a non-zero value for $\mu_\nu$~\cite{Voloshin}, or
employing a spin suppression mechanism to keep $m_\nu$
small~\cite{Barr}.
Neutrino magnetic moments are reviewed in~\cite{Fukugita,boris,Wong}, and 
recent work can be found in~\cite{McLaughlin}.

When $\Lambda$ is not substantially larger than $v$, the contribution
to $\delta m_\nu$ from higher dimension operators can be important,
and their renormalization due to operators responsible for the
neutrino magnetic moment can be computed in a model-independent way.
As we discuss below, dimension six operators are the lowest that contribute.
We shall now outline 
this calculation and the
resulting constraints on $\mu_\nu$. Specifically, we find that
\begin{equation}
\label{eq:massmurel}
\frac{|\mu_\nu|}{\mu_B} = 
\frac{G_F \, m_e }{\sqrt{2}\pi \alpha }\ 
\left[ \frac{\delta m_\nu}{  \alpha \, \ln ( \Lambda/v) } \right]  
\ \frac{32\pi\sin^4\theta_W}{9 \,  |f|} \ \ \ , 
\end{equation}
where $\theta_W$ is the weak mixing angle,
\begin{equation}
f=\left(1-r \right)-\frac{2}{3} r \tan^2\theta_W
-\frac{1}{3}\left(1+r\right)\tan^4\theta_W\ \ \ ,
\end{equation}
and $r=C_{-}/C_{+}$ is a ratio of effective operator coefficients
defined at the scale $\Lambda$ (see below) that one expects to be of
order unity.  Again taking $\Lambda \sim \ 1$ TeV, $\delta m_\nu \alt
1 {\rm eV}$, and setting $r\sim 1$, we find
that $\mu_\nu$ (for any mass eigenstate) should be smaller in
magnitude than $\sim 10^{-14}\mu_B$.

To arrive at these conclusions, we consider an effective theory
containing Dirac fermions, scalars, and gauge bosons that is valid
below the scale $\Lambda$ and that respects the SU(2)$_L\times$
U(1)$_Y$ symmetry of the SM. We also impose lepton number conservation.
In this effective theory, the
right-handed components of the neutrino have zero hypercharge ($Y$)
and weak isospin. The effective Lagrangian involving the $\nu_R$,
left-handed lepton isodoublet $L$, and Higgs doublet $\phi$ obtained
by integrating out physics above the scale $\Lambda$ is given by 
\begin{equation}
\label{eq:leff}
{\cal L}_{\rm eff} = \sum_{n,j} {C^n_j(\mu)\over \Lambda^{n-4}}\ {\cal O}^{(n)}_j(\mu)\ +\ \ {\rm h.c.}\ \,
\end{equation}
where the $n\geq 4$ denotes the operator dimension, $j$ runs over all
independent operators of a given dimension, and $\mu$ is the
renormalization scale. For simplicity, we do not write down the $n=4$
operators appearing in the SM Lagrangian or the Dirac Lagrangian for
the $\nu_R$. At $n=4$, a neutrino mass would arise from the operator $
{\cal O}^{(4)}_1 = {\bar L}{\tilde\phi}\nu_R$, where
${\tilde\phi}=i\tau_2\phi^\ast$. We also omit explicit flavor indices
on the $L$ and $\nu_R$ fields. After spontaneous symmetry breaking
(SSB) at the weak scale,
\begin{equation}
\phi\rightarrow\left(
\begin{array}{c}
0 \\
v/\sqrt{2}
\end{array}
\right)
\end{equation}
so that $C^4_1{\cal O}^{(4)}_1\to -m_\nu {\bar\nu}_L\nu_R$ with
$m_\nu= -C^4_1v/\sqrt{2}$. Consistency with the present information on
the scale of $m_\nu$ requires that $|C^4_1|\lsim 5\times 10^{-12}$.

A neutrino magnetic moment coupling would be generated by
gauge-invariant, dimension six operators that couple the matter fields
to the SU(2)$_L$ and U(1)$_Y$ gauge fields $W_\mu^a$ and $B_\mu$,
respectively. Above the scale $v$, these operators will mix under
renormalization with other $n=6$ operators that contain the $L$,
$\nu_R$, and $\phi$ and that generate neutrino mass terms after
SSB. For this purpose, the basis of independent $n=6$ operators that
close under renormalization is given by 
\bea \nonumber 
{\cal O}^{(6)}_1 & = & g_1{\bar L}{\tilde \phi}\sigma^{\mu\nu}\nu_R B_{\mu\nu} \\
\label{eq:ops}
{\cal O}^{(6)}_2 & = & g_2 {\bar L}\tau^a {\tilde \phi} 
\sigma^{\mu\nu}\nu_R W_{\mu\nu}^a \\ \nonumber 
{\cal O}^{(6)}_3 & = & {\bar L}{\tilde \phi}\nu_R \left(\phi^\dag\phi\right) \ \ \ , 
\eea
where $B_{\mu\nu}=\partial_\mu B_\nu-\partial_\nu B_\mu$ and
$W_{\mu\nu}^a=\partial_\mu W_\nu^a-\partial_\nu
W_\mu^a-g_2\epsilon_{abc}W_\mu^b W_\nu^c$ are the U(1)$_Y$ and
SU(2)$_L$ field strength tensors, respectively, and $g_1$ and $g_2$
are the corresponding couplings. After SSB one has
\bea {\cal
O}^{(6)}_1 & \to & \frac{v}{\sqrt{2}} g_1{\bar \nu}_L
\sigma^{\mu\nu}\nu_R B_{\mu\nu} \\ {\cal O}^{(6)}_2 &\to &
g_2\frac{v}{\sqrt{2}} {\bar \nu}_L \sigma^{\mu\nu}\nu_R W_{\mu\nu}^3
+\cdots \ \ \ .  \eea 
Using $g_2\sin\theta_W=g_1\cos\theta_W=e$, it is straightforward to
see that the combination $C^6_1{\cal O}^{(6)}_1+C^6_2 {\cal
O}^{(6)}_2$ appearing in ${\cal L}_{\rm eff}$ contains the magnetic
moment operator
\begin{equation}
-{\mu_\nu \over 4} {\bar\nu} \sigma^{\mu\nu}\nu F_{\mu\nu}
\end{equation}
where $F_{\mu\nu}$ is the photon field strength tensor and
\be
\label{eq:munu1}
\frac{\mu_\nu}{\mu_B}=
-4\sqrt{2}\left( \frac{m_e v}{\Lambda^2}\right) \left[C^6_1(v)+C^6_2(v)\right]\ \ \ .
\ee
Similarly, the operator ${\cal O}^{(6)}_3$ generates a contribution to the neutrino mass  
\be
\label{eq:deltamnu1}
\delta m_\nu = -C^6_3(v) \frac{v^3}{2\sqrt{2}\Lambda^2} \ \ \ .
\ee

Other $n=6$ operators that one can write down are either related to
those in Eqs. (\ref{eq:ops}) by the equations of motion or do not
couple to $F_{\mu\nu}$ after SSB. It is instructive to consider a few
illustrative examples. In particular, consider the following three
operators:
\bea
\label{eq:otherops}
{\cal O}^{(6)}_4 & = & {\bar L} {\overleftarrow D}_\mu^\dag {\overleftarrow D}^{\mu\ \dag}{ \tilde\phi} \nu_R \\
{\cal O}^{(6)}_5 & = & {\bar L} {\overleftarrow D}_\mu^\dag{\tilde\phi}\partial^\mu \nu_R \\
{\cal O}^{(6)}_6 & = & {\bar L} \tau^a {\tilde\phi} \nu_R \left(\phi^\dag\tau^a\phi\right)\ \ \ ,
\eea
where $D_\mu = \partial_\mu +ig_2 \tau^a W_\mu^a/2+ig_1 Y B_\mu/2$ and 
where the sum over $a=1,2,3$ in ${\cal O}^{(6)}_6$ is implied.
We may express  ${\cal O}^{(6)}_4$ in terms of ${\cal O}^{(6)}_{1,2}$ by first noting that 
\be
0={\bar L} {\overleftarrow {\not\!\!  D}}^\dag {\overleftarrow {\not\!\!  D}}^\dag {\tilde\phi}\nu_R
\ee
since ${\bar L} {\overleftarrow {\not\! \! D}}^\dag=0={\not\!\! D} L$
by the equation of motion for $L$. Then using
$\gamma_\mu\gamma_\nu=g_{\mu\nu}-i\sigma_{\mu\nu}$ we have
\be
0={\bar L} {\overleftarrow D}_\mu^\dag {\overleftarrow D}^{\mu \dag}{\tilde\phi} \nu_R
-\frac{i}{2} {\bar L}[{\overleftarrow D}_\mu^\dag, {\overleftarrow D}_\nu^\dag]\sigma^{\mu\nu} {\tilde\phi}\nu_R\ \ \ .
\ee
Working out the commutator $[{\overleftarrow D}_\mu^\dag,
{\overleftarrow D}_\nu^\dag]$ in terms of $B_{\mu\nu}$ and
$W_{\mu\nu}^a$ gives
\be
{\bar L} {\overleftarrow D}_\mu^\dag {\overleftarrow D}^{\mu \dag}{\tilde\phi} \nu_R = 
-\frac{1}{4}\left[Y_L{\cal O}^{(6)}_1+{\cal O}^{(6)}_2\right]\ \ \ .
\ee
where $Y_L=-1$ is the lepton doublet hypercharge.

\begin{figure}[t!]
\includegraphics[height=2.2cm]{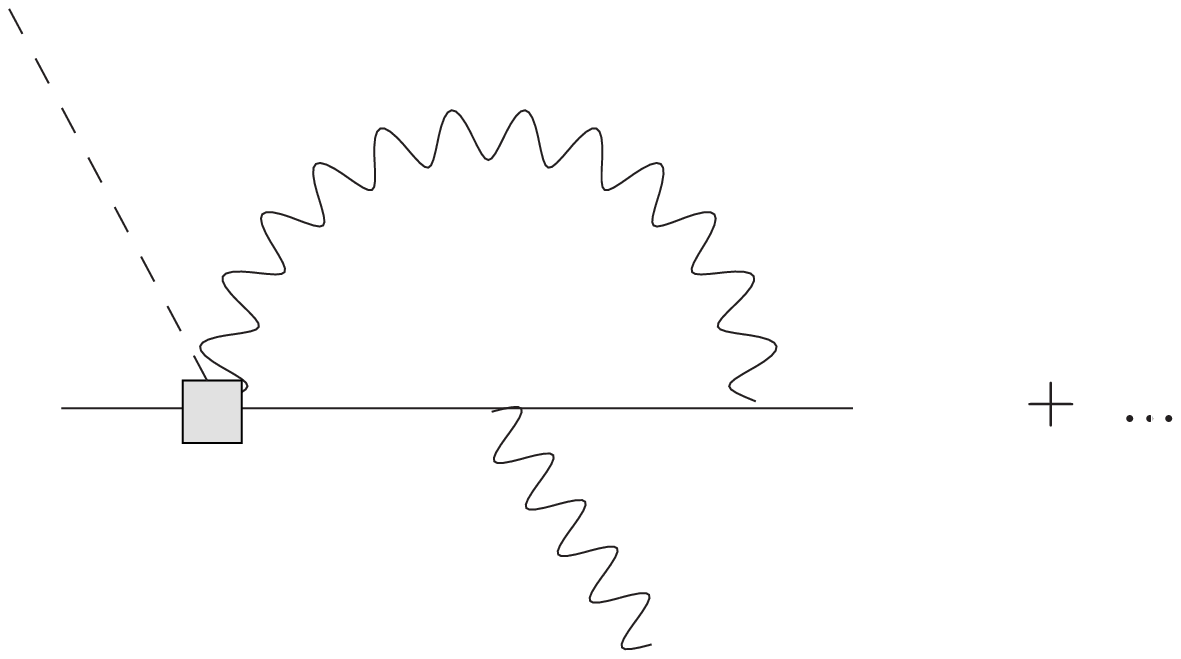}
\caption{
Self-renormalization of ${\cal O}_{1,2}^{(6)}$, denoted by the
shaded box. Solid, dashed, and wavy lines indicate leptons, Higgs, and
gauge bosons, respectively.}
\label{fig:fig1}
\includegraphics[height=2.0cm]{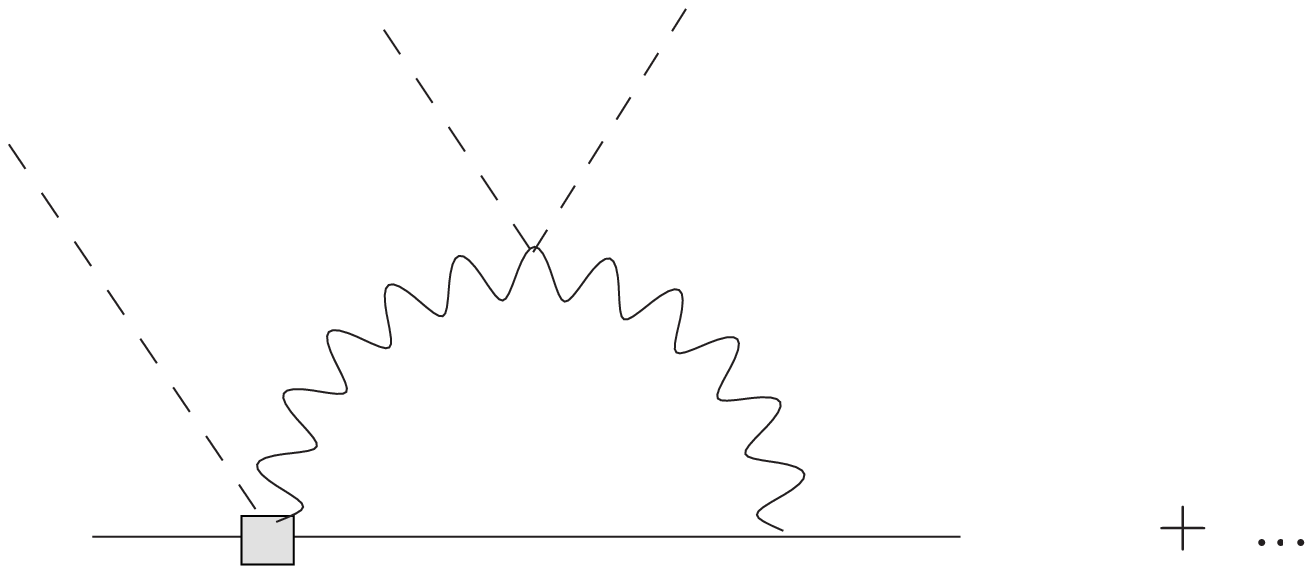}
\caption{ 
Renormalization of ${\cal O}^{(6)}_3$ due to insertions of
${\cal O}_{1,2}^{(6)}$.}
\label{fig:fig2}
\includegraphics[width=7.0cm]{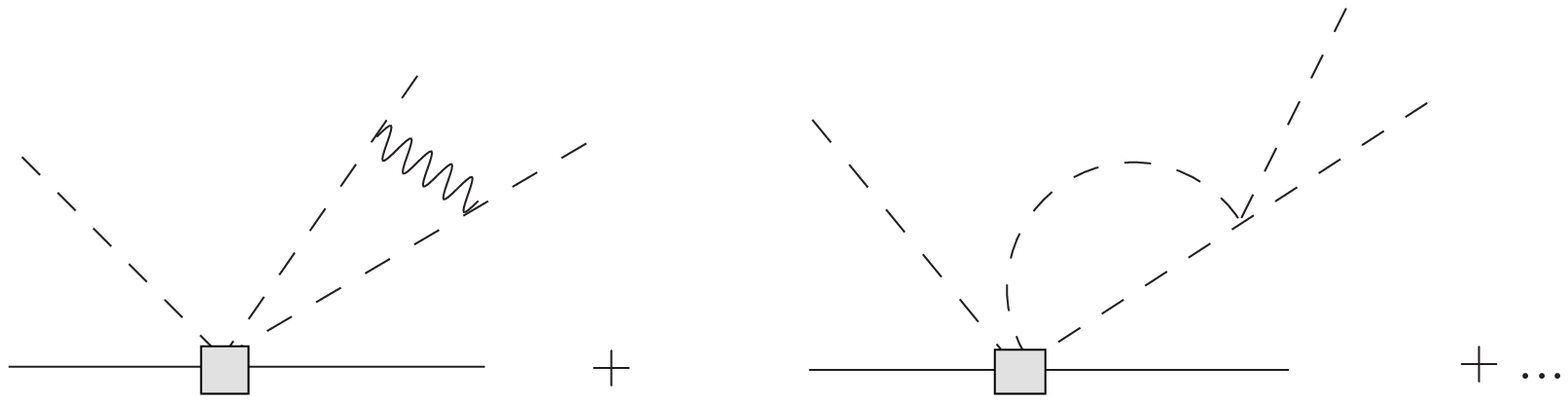}
\caption{
Self-renormalization of ${\cal O}_{3}^{(6)}$.}
\label{fig:fig3}
\end{figure}

In the case of ${\cal O}^{(6)}_5$, the component involving ${\bar
\nu}_L {\overleftarrow D}_\mu^\dag \phi^{0 \ast}\partial^\mu\nu_R$
contains only the combination $g_2 W_\mu^3-g_1 B_\mu\propto Z_\mu$
since the SM Lagrangian for the neutrino contains no coupling to the
photon.  Moreover, since ${\cal O}^{(6)}_5$ contains a derivative
acting on $\nu_R$ and $\nu_R$ has no gauge interactions, it does not
mix with ${\cal O}^{(6)}_{1-3}$ under renormalization.  Finally, one
can show that ${\cal O}^{(6)}_6=-{\cal O}^{(6)}_3$ using the
identity $\tau^a_{ij} \tau^a_{kl}=
2\delta_{il}\delta_{jk}-\delta_{ij}\delta_{kl}$. Other operators that
contain derivatives acting on the $\phi$ may be related to ${\cal
O}^{(6)}_{1,2}$ using integration by parts and the foregoing
arguments.

The one-loop renormalization of the ${\cal O}^{(6)}_{1-3}$ is obtained
by computing Feynman diagrams of Figs. 1-3, where only representative
examples of the full set of graphs are shown. The graphs of
Fig.~\ref{fig:fig1} involve renormalization of ${\cal O}^{(6)}_{1,2}$,
where the shaded box indicates insertions of the tree-level operator.
Graphs of the type shown in Fig.~\ref{fig:fig2} give renormalization
of ${\cal O}^{(6)}_3$ by insertions of the ${\cal O}^{(6)}_{1,2}$. At
this order, there are no insertions of ${\cal O}^{(6)}_3$ that
renormalize the ${\cal O}^{(6)}_{1,2}$. Graphs leading to
self-renormalization of ${\cal O}^{(6)}_3$ generated by gauge and
$\lambda(\phi^\dag\phi)^2$ couplings are illustrated in
Fig.~\ref{fig:fig3}. For the diagrams involving internal gauge boson
lines, we use the background field gauge\cite{Abbott:1980hw}, which
allows us to obtain gauge-invariant results in a straightforward
manner.  Throughout, we use
dimensional regularization, working in $d=4-2\epsilon$
dimensions, and introduce the 
renormalization scale
$\mu$. Due to operator mixing, the renormalized operators ${\cal
O}^{(6)}_{j R}$ can be expressed in terms of the un-renormalized
operators ${\cal O}^{(6)}_j$ 
via
\be
{\cal O}^{(6)}_{jR} = \sum_k Z^{-1}_{jk} Z_L^{1/2} Z_\phi^{n_\phi/2} {\cal O}^{(6)}_{k} \ \ \ , 
\ee
where $Z_L^{1/2}$ and $Z_\phi^{1/2}$ are wavefunction
renormalization constants for the $L$ and $\phi$ respectively and
where $n_\phi=1$ (3) is the number of $\phi$ fields appearing in
${\cal O}^{(6)}_{1,2}$ (${\cal O}^{(6)}_{3}$). In the minimal
subtraction scheme that we adopt here, the products of renormalization
constants $Z^{-1}_{jk} Z_L^{1/2} Z_\phi^{n_\phi/2}$ simply remove the
$1/\epsilon$ terms arising from the loop graphs.

The renormalized operators ${\cal O}^{(6)}_{jR}$ are dependent on the
scale $\mu$ since the bare operators
\be
{\cal O}^{(6)}_{j0}= Z_L^{1/2} Z_{\phi}^{n_\phi/2} {\cal O}^{(6)}_{j} 
= \sum_k Z_{jk}{\cal O}^{(6)}_{k R}
\ee
must be $\mu$-independent. The $\mu$-dependence of the $C_i^6(\mu)$ is
such that the renormalized effective Lagrangian ${\cal L}_{\rm eff}^R$
does not depend on the renormalization scale.  In order to obtain
Eq. (\ref{eq:massmurel}) we require the value of the $C^6_i(\mu)$ at
the scale $\mu=v$, below which the $Z$ and $W^\pm$ are integrated out
of the effective theory and only the photon contributes to operator
renormalization. Since $Q_\nu=0$, the latter occurs at higher order in
$\alpha/4\pi$ than considered here. The value of the $C^6_i(v)$ are
determined by the renormalization group equation (RGE) that follows
from the requirement that ${\cal L}_{\rm eff}^R$ be $\mu$-independent:
\be
\label{eq:rge}
\mu\frac{d}{d\mu}{\cal L}_{\rm eff}^R = 0\ \Rightarrow \ 
\mu \frac{d}{d \mu}C^6_j +\sum_kC^6_k\ \gamma_{kj}=0 \ \ \ , 
\ee
where the anomalous dimension matrix is defined by
\be
\gamma_{kj} = \sum_{\ell} \left( \mu\frac{d}{d\mu} Z_{k\ell }^{-1}\right)  Z_{\ell j} \ \ \ .
\ee
We find 
\begin{widetext}
\be
\label{eq:anomdim}
\gamma_{jk} =
\left(
\begin{array}{ccc}
-\frac{3}{16\pi}(\alpha_1-3\alpha_2) & \frac{3}{8\pi}\alpha_1 & -6\alpha_1(\alpha_1+\alpha_2)\\
\frac{9}{8\pi}\alpha_2& \frac{3}{16\pi}(\alpha_1-3\alpha_2)& 6\alpha_2(\alpha_1+3\alpha_2)\\
0 & 0 & \frac{9}{16\pi}(\alpha_1+3\alpha_2)-\frac{5}{4\pi^2}\lambda
\end{array}
\right)\ \ \ ,
\ee
\end{widetext}
where the $\alpha_i=g_i^2/4\pi$ and
$V(\phi)=\lambda[(\phi^\dag\phi)-v^2/2]^2$.

Using the known $\beta$-functions that govern the $\mu$-dependence of
the $g_i$ and the anomalous dimension matrix in Eq. (\ref{eq:anomdim})
we numerically solve the RGE (\ref{eq:rge}) for the $C^6_i(\mu)$. In
doing so, we find that the $\mu$-dependence of the $g_i$ has a
negligible impact on the overall solution. Neglecting the
$\mu$-dependence of the $g_i$ then allows us to obtain an analytic
solution. In this approximation we find that the combinations of
constants $C_{+}(\mu)\equiv C_1^6(\mu)+C_2^6(\mu)$ and
$\tilde{C}(\mu)\equiv \alpha_1C_1^6(\mu)-3\alpha_2C_2^6(\mu)$ evolve independently. Since
$\mu_\nu$ is proportional to $C_{+}(v)$, the presence of a non-zero
neutrino magnetic moment at low energy requires the physics beyond the
SM to have generated a non-vanishing $C_{+}(\Lambda)$. It is then
straightforward to obtain $C_3^6(\mu)$ which depends on all three of
the $C_i^6(\Lambda)$. Retaining only the leading logarithms rather
than the full resummation provided by the RGE and defining $C_{-}(\mu)
= C^6_1(\mu) - C^6_2(\mu)$ we find
\bea
\label{eq:cimu}
C_+(\mu)&  = & C_+(\Lambda)\left[1-{\tilde\gamma} \ln\frac{\mu}{\Lambda}\right]\\
\nonumber
\tilde{C} (\mu) & = &\tilde{C}(\Lambda)\left[1+ {\tilde\gamma} \ln\frac{\mu}{\Lambda}\right]\\
\nonumber
C^6_3(\mu) & = & C^6_3(\Lambda)\left[1-\gamma_{33}\ln\frac{\mu}{\Lambda}\right]\\
\nonumber
&-& \left[C_{+}(\Lambda)\gamma_{+}+C_{-}(\Lambda)\gamma_{-}\right]\ln\frac{\mu}{\Lambda}\ \ \ ,
\eea
where $\gamma_\pm=(\gamma_{13}\pm\gamma_{23})/2$ and ${\tilde\gamma}=3(\alpha_1+3\alpha_2)/16\pi$.

Using Eqs. ({\ref{eq:munu1},\ref{eq:deltamnu1}) allows us to relate
$\mu_\nu$ to the corresponding neutrino mass matrix element in terms
of
$C_{\pm}(\Lambda)$ and $C^6_3(\Lambda)$
\be
\label{eq:munurel}
\delta m_\nu=\frac{v^2}{16 m_e}\ \frac{C^6_3(v)}{C_+(v)}\ \frac{\mu_\nu}{\mu_B}\ \ \ \ ,
\ee
with $C_+(v)$ and $C^6_3(v)$ given approximately by
Eqs. (\ref{eq:cimu}). To obtain a natural upper bound on $\mu_\nu$, we
assume first that $C^6_3(\Lambda)=0$ so that $\delta m_\nu$ is
generated entirely by radiative corrections involving insertions of
${\cal O}^{(6)}_{1,2}$. Doing so in Eq. (\ref{eq:munurel}) and solving
for $\mu_\nu/\mu_B$ leads directly to Eq. (\ref{eq:massmurel}). To
arrive at a numerical estimate of this bound, we substitute $\Lambda =
1$ TeV into the logarithms appearing in Eq. (\ref{eq:massmurel}) and
obtain
\be
 \label{eq:massbound}
 \frac{|\mu_\nu|}{\mu_B} \lsim 8\times 10^{-15}\times \left(\frac{\delta m_\nu}{1\ {\rm eV}}\right) \frac{1}{|f|} \ \ \ .
 \ee
 
It is interesting to consider the bound for the special case that only
the magnetic moment operator is generated at the scale $\Lambda$, {\em
i.e.}, $C_{+}(\Lambda)\not= 0$ and $C_{-}=0$, with $f \simeq 1$.  For
this case, considering a nearly degenerate neutrino spectrum with
masses $\sim 1$ eV leads to the $|\mu_\nu| \lsim 8\times
10^{-15}\mu_B$ -- a limit that is two orders of magnitude stronger
than the astrophysical bound~\cite{raffelt} and $10^4$ stronger than
those obtained from solar and reactor neutrinos. For a hierarchical
neutrino mass spectrum, the bound would be even more stringent.

The discovery of a Dirac neutrino magnetic moment having a magnitude
comparable to, or just below, the present experimental limits would
imply considerable fine-tuning in order to maintain consistency with
the scale of neutrino mass. Such fine-tuning could occur through
cancellations between the $C_{+}(\Lambda)$, $C_{-}(\Lambda)$, and
$C^6_3(\Lambda)$ terms in Eq.~(\ref{eq:cimu}).  While it is in
principle possible to construct a model that displays such
fine-tuning, the generic situation implies substantially smaller
magnetic moments for Dirac neutrinos than are presently accessible
through observation.

The limits one may obtain on transition magnetic moments of Majorana
neutrinos are substantially weaker than those for the Dirac
moment. Because the transition magnetic moment $\mu_\nu^{ij}$ is
antisymmetric in the flavor labels $i$, $j$, while the mass matrix is
symmetric, $\delta m_\nu$ must be higher order in $\mu_\nu^{ij}$ or
involve insertions of the Yukawa couplings.

\acknowledgments
This work was supported in part under U.S. Department of Energy
contracts \# DE-FG02-05ER41361 and DE-FG03-92ER40701, and NSF 
grant PHY-0071856.

\end{document}